\documentclass[conference, 10pt]{IEEEtran}
\IEEEoverridecommandlockouts
\usepackage{cite}
\usepackage{amsmath,amssymb,amsfonts}
\usepackage{algorithmic}
\usepackage{graphicx}
\usepackage{textcomp}
\usepackage{xcolor}
\usepackage[normalem]{ulem}
\def\BibTeX{{\rm B\kern-.05em{\sc i\kern-.025em b}\kern-.08em
    T\kern-.1667em\lower.7ex\hbox{E}\kern-.125emX}}
    
\usepackage{amsmath,amssymb}
\usepackage{mathtools} 
\usepackage{xspace}

\newcommand{\Secref}[1]{Section~\ref{#1}}
\newcommand{\Figref}[1]{Fig.~\ref{#1}}

\newcommand{\herm}{^\text{H}}

\newcommand{\bw}{\mathbf{w}}

\newcommand{\by}{\mathbf{y}}

\newcommand{\bphi}{\boldsymbol{\phi}}

\newcommand{\tc}{\tau_{\textsc{c}}}

\newcommand{\boundellipse}[3]
{(#1) ellipse [x radius=#2,y radius=#3]
}

\DeclareMathOperator{\var}{\mathsf{Var}}

\DeclareMathOperator*{\argmax}{arg\,max}

\newcommand{\EX}[1]{\mathsf{E}\left\{{#1}\right\}}

\newcommand{\varx}[1]{\var\left\{{#1}\right\}}

\newcommand{\C}{\mathbb{C}}

\newcommand{\Pp}{\rho_{\mathrm{p}}}

\newcommand{\norm}[1]{{ \left\Vert #1 \right\Vert }}

\newcommand{\tp}{\tau_{\textsc{p}}}

\makeatletter
\def\@setsize#1#2#3#4{
    \@nomath#1
    \let\@currsize#1
    \baselineskip #2
    \baselineskip \baselinestretch\baselineskip
    \parskip \baselinestretch\parskip
    \setbox\strutbox \hbox{
        \vrule height.7\baselineskip
            depth.3\baselineskip
            width\z@}
    \skip\footins \baselinestretch\skip\footins
    \normalbaselineskip\baselineskip#3#4}
\makeatother

\makeatletter
\newcommand{\setstretch}[1]{
    \def\baselinestretch{#1}%
    \@currsize
    }
\makeatother



\begin{document}

\begin{figure*}[t!]
\normalsize
Paper accepted for presentation in IEEE SPAWC 2020 - 21st IEEE International Workshop on Signal Processing Advances in Wireless Communications. 

\

\textcopyright~2020 IEEE. Personal use of this material is permitted.  Permission from IEEE must be obtained for all other uses, in any current or future media, including reprinting/republishing this material for advertising or promotional purposes, creating new collective works, for resale or redistribution to servers or lists, or reuse of any copyrighted component of this work in other works.
\vspace{17cm}
\end{figure*}

\title{Self-Learning Detector for the Cell-Free Massive MIMO Uplink: The Line-of-Sight Case
\thanks{This paper was supported in part by the Swedish Research Council (VR), and in part by ELLIIT.}
}


\author{\IEEEauthorblockN{Giovanni Interdonato$^{*}$, P{\aa}l Frenger$^{\dagger}$ and Erik G. Larsson$^*$}
\IEEEauthorblockA{$^*$Dept. of Electrical Engineering (ISY), Link\"oping University, Link\"oping, Sweden\\$^\dagger$Ericsson Research, Link\"oping, Sweden\\
\{giovanni.interdonato, erik.g.larsson\}@liu.se, pal.frenger@ericsson.com}}

\maketitle

\begin{abstract}
The precoding in cell-free massive multiple-input multiple-output (MIMO) technology relies on accurate knowledge of channel responses between users (UEs) and access points (APs).  Obtaining high-quality channel estimates in turn requires the path losses between pairs of UEs and APs to be known. These path losses may change rapidly especially in line-of-sight environments with moving blocking objects.  A difficulty in the estimation of path losses is pilot contamination, that is, simultaneously transmitted pilots from different UEs that may add up destructively or constructively by chance, seriously affecting the estimation quality (and hence the eventual performance).  A method for estimation of path losses, along with an accompanying pilot transmission scheme, is proposed that works for both Rayleigh fading and line-of-sight channels and that significantly improves performance over baseline state-of-the-art. The salient feature of the pilot transmission scheme is that pilots are structurally phase-rotated over different coherence blocks (according to a pre-determined function known to all parties), in order to create an effective statistical distribution of the received pilot signal that can be efficiently exploited by the proposed estimation algorithm.
\end{abstract}


\section{Introduction}
In cell-free massive MIMO (multiple-input multiple-output)~\cite{Ngo2017b,Interdonato2018a,Zhang2019}, potentially large numbers of access points (APs) are distributed over a wide geographical area.
These APs simultaneously serve many users (UEs) through coherent precoding, in time-division duplex (TDD) mode.
In its canonical form, cell-free massive MIMO relies on uplink pilots transmitted by the UEs in order to estimate all UE-to-AP uplink channel responses. 
These estimates are then used to aid the uplink data decoding, and by virtue of reciprocity of propagation, subsequently for the downlink precoding. 

Methods typically used for channel estimation in the literature are based on Bayesian minimum-mean square error (MMSE) estimation~\cite{Kay1993a}.  
MMSE estimation requires a priori assumptions to be made on the statistics of the channel responses. For instance, let $g_{mik}$ be the scalar channel response between a single-antenna AP $m$ and a single-antenna UE $k$, in the $i$-th coherence block.
The standard assumption is that $\{ g_{mik} \}$ are statistically independent, and $g_{mik}$ is zero-mean complex Gaussian, with known variance $\beta_{mk}=\varx{g_{mik}}$ that represents the average (over multiple coherence blocks) channel path loss, and includes large-scale shadowing effects~\cite{Ngo2017b,Nayebi2017a,Buzzi2019c,Bashar2019,Liu2020}. 
This assumption corresponds to Rayleigh fading channels. 

There is no known way around the assumption that the path losses are known, other than to use special training data to estimate them, which requires the expense of significant extra dedicated resources~\cite{Bjornson2016c,Kocharlakota2019}. 
These resources may be simply unavailable in applications that require ultra-low latency.
A complication in this estimation is that pilots are typically reused (because of  finite channel coherence), which introduces ``pilot contamination'' interference that is hard to resolve without a priori information~\cite{Marzetta2016a,Jose2011b,Yin2013a,Haghighatshoar2017a,Bjornson2018}.
More  importantly, the entire notion of a Gaussian prior on $\{ g_{mik} \}$ requires stationarity, that is, constancy of $\{ \beta_{mk} \}$ over time and frequency, an assumption that is likely to be violated in practice. 
For example, a fast moving blocking object (or the UE itself moving behind a blocking object) may abruptly change the path loss, especially at higher carrier frequencies. 
In conclusion, the stationarity assumption and the associated requirement of prior knowledge of $\{ \beta_{mk} \}$ can be hampering in a practical implementation.

It might be tempting to use other algorithms than MMSE for the estimation of $\{ g_{mik} \}$, that do not require any prior assumptions, such as the non-Bayesian ``least-squares'' estimator. 
However, that results in very poor performance~\cite{Bjornson2016c,Ozdogan2019} unless appropriate post-processing of the received data is used, post-processing which in turn requires knowledge of $\{ \beta_{mk} \}$.   
The reason is that $\{ \beta_{mk} \}$ contain a significant amount of information---in fact, the information encoded in $\{ \beta_{mk} \}$ can be interpreted as  a priori information on the UE locations. The larger a value of $\beta_{mk}$ is, the closer a UE $k$ to an AP $m$ is.

\textbf{Contribution:} We propose a solution to estimate the channel responses and the path losses in the presence of pilot contamination, along with an accompanying phase-rotation pilot transmission scheme.  The solution is particularly useful in line-of-sight operation, which is likely to be a most common operating condition for cell-free massive MIMO systems.\footnote{A practical implementation of cell-free massive MIMO denoted \textit{radio stripes} is proposed in~\cite{Interdonato2018a,patentRadioStripeUS}. Radio stripes aim at enabling invisible, low-cost deployment of large number of distributed APs in the vicinity of the UEs. 
This is achieved by serial integration of distributed transmitting and receiving components into the same cable that also provide fronthaul communication and power. 
Since the UEs are surrounded by many APs, it is reasonable to assume that some APs are in line-of-sight to each UE.} 
The solution is also especially useful in scenarios where $(i)$ the stationarity assumption does not hold (i.e., fast-changing blocking conditions), $(ii)$ at higher carrier frequencies, $(iii)$ where latency is a concern, $(iv)$ where the allocated bandwidth is small, e.g., certain mMTC (massive machine-type communications) and IoT (Internet-of-Things) scenarios. Unlike prior-art schemes~\cite{Bjornson2016c,Kocharlakota2019}, our proposed scheme does not require additional dedicated resources for estimating the path losses. As in~\cite{Neumann2018}, the path losses are estimated in the same resources employed for the channel response estimation. Each AP self-learns the path losses during the uplink training, wherein an a-priori-assumption-free channel estimate is performed. 
However, the scheme proposed in~\cite{Neumann2018} requires \textcolor{black}{to reallocate the pilots according to proper patterns in each coherence block in order to reconstruct the channel covariance matrices (i.e., the joint pilot allocation matrix must be full rank and known to all the parties). Hence, the analysis in~\cite{Neumann2018} is entirely different from the analysis herein proposed, and it is also limited to independent Rayleigh fading channels}.
To the best of author's knowledge, there are no studies on path losses (i.e., covariance matrix) estimation considering the line-of-sight channel model. \textcolor{black}{The analysis in this work focuses on estimating the path losses of two UEs sharing the same pilot over multiple coherence blocks. (Path loss estimation through orthogonal resources is trivial and therefore omitted.) An extension to multiple co-pilot UEs remains a possible topic for future work.}   

\section{System Model}
\label{sec:system-model}

$K$ single-antenna UEs are served through coherent precoding by $M$ service antennas. These $M$ service antennas are deployed on APs. An AP may have a single antenna, or (small) arrays of antennas; the precise arrangement is substantially immaterial for the modeling and only affects the eventual performance. For simplicity, we assume single-antenna APs and single-antenna UEs in this discussion. Generalization to multi-antenna APs and UEs is straightforward.

There is full coherent cooperation among all service $M$ APs. 
The channel coherence block consists of $\tc$ samples of which $\tp$ are used for uplink pilots.
A set of $\tp$ pre-determined orthonormal sequences ($\tp$-length vectors) 
are used as pilots.
The case of interest is when $K > \tp$, so that reuse of pilots among different UEs is inevitable. Only the uplink is of concern here, and the transmission in $I$ coherence blocks is considered. 
\textcolor{black}{We assume that the path losses are constant within the $I$ coherence blocks.}
In a low-latency application, these coherence blocks would consist of  groups of subcarriers of a single OFDM (orthogonal frequency-division multiplexing) symbol in time, though nothing precludes the $I$ blocks to span over multiple OFDM symbols in principle.
\subsection{Uplink training}
For the purpose of channel estimation, the $K$ UEs transmit uplink pilots in each coherence block. Let $p_{ik}$ be the index of the pilot sequence used by UE $k$ in the $i$-th coherence block. 
We denote $\sqrt{\tp\Pp}\bphi_{p_{ik}} \in \C^{\tp}$ as the pilot sequence sent by the $k$-th UE, $k=1,...,K$, where $\norm{\bphi_{p_{ik}}}=1$, and $\Pp$ is a constant that has the interpretation of pilot signal-to-noise ratio (SNR). We assume that any two pilot sequences are either identical or mutually orthonormal, that is $\bphi_{p_{ik}}\herm \bphi_{{p_{ik'}}}$ is either equal to 1, if $\bphi_{p_{ik}}\!=\!\bphi_{{p_{ik'}}}$, or 0, otherwise.
The pilot signal received at AP $m$ is a linear superposition of $K$ pilots:
\begin{equation}
\label{eq:uplinkpilot}
\by_{mi} = \sqrt{\Pp \tp} \sum\nolimits^K_{k=1} g_{mik} \bphi_{p_{ik}} + \bw_{mi},
\end{equation}
where $\bw_{mi}\in \C^{\tp}$ is the receiver noise vector whose elements are i.i.d. $\mathcal{CN}(0,1)$. 
AP $m$ performs ``de-spreading'' in the standard manner by projecting this received pilot signal onto the orthonormal pilot vectors $\{ \bphi_1,\ldots,\bphi_{\tp} \}$. This de-spreading results in $I\tp$ random variables,
\begin{align}
y_{miq} &= \frac{1}{\sqrt{\tp}} \bphi_q\herm \by_{mi} =\sqrt{\Pp} \sum_{k: p_{ik}=q} g_{mik} + \frac{w_{miq}}{\sqrt{\tp}},
\end{align}
$i=1,\ldots,I; \quad q=1,\ldots,\tp$.
Each variable, $y_{miq}$, contains the received pilot at the $m$-th AP in the $i$-th coherence block projected onto the $q$-th pilot sequence.
The sum is over those UEs that use the $q$-th pilot sequence, and this summation arises because of the pilot reuse.\footnote{If each UE would had a unique pilot sequence, then $\sum_{k: p_{ik}=q} g_{mik}$ would reduce to $g_{mik'}$ where $k'$ is the index of the UE that uses pilot $q$.}
The terms $\{ w_{miq} = \bphi_q\herm \bw_{mi} \}$ contain estimation noise and are mutually independent $\mathcal{CN}(0,1)$.
\subsection{Channel Estimation}
The canonical assumption made in all related papers the authors are aware of assumes that a priori, $\{ g_{mik} \}$ are statistically independent, $g_{mik} \sim \mathcal{CN}(0,\beta_{mk})$, where $\{ \beta_{mk} \}$ are known, from which the MMSE estimate easily follows~\cite{Ngo2017b}:
\begin{align}\label{eq:mmse}
\hat g_{mik}  &= \EX{ g_{mik}  | \{ y_{mi1} , \ldots, y_{mi\tp} \} } = \EX{ g_{mik}  | y_{mip_{ik}} } \nonumber \\ 
&=  \frac{\sqrt{\Pp} \tp \beta_{mk} } { \Pp\tp\sum_{k': p_{ik'}=p_{ik} }^K \beta_{mk'} + 1} y_{mik}.
\end{align}
The mean-square of these channel estimates is,
\begin{align}
\gamma_{mik}  = \EX{  |\hat g_{mik}|^2} = \frac{\Pp \tp \beta^2_{mk} } { \Pp\tp\sum_{k': p_{ik'}=p_{ik}}^K \beta_{mk'} + 1}  ,
\end{align}
and represents a quality measure of the estimate: it always holds   $\gamma_{mik}\le \beta_{mk}$, and the closer
$\gamma_{mik}$ is to $ \beta_{mk}$, the better is the estimate (the less is the effect of pilot contamination and measurement noise).
The channel estimates $\{ \hat g_{mik} \}$ in (\ref{eq:mmse}) are optimal if $g_{mik}~\sim~\mathcal{CN}(0,\beta_{mk})$, that is independent Rayleigh fading, and sub-optimal otherwise.
They are typically useful also for other fading distributions (e.g., Ricean, line-of-sight) as long as $\beta_{mk}$ has the meaning of ``average strength'' (mean-square value) of $g_{mik}$.

\section{Self-Learning Detector}
\label{sec:two-user-system}
For given $q$, the variables  $\{y_{miq} \}$ constitute a sufficient statistic for the estimation of all $\{ g_{mik} \}$ for which $p_{ik}=q$.
However, without the use of additional prior knowledge, estimates based on $\{ y_{miq} \}$ are typically meaningless. 
To understand why, consider a scenario with $K=2$ UEs that both use the first pilot ($q=1$) in every coherence block. Then
\begin{align}
y_{mi1} = \sqrt{\Pp}( g_{mi1} + g_{mi2} ) + \frac{1}{\sqrt{\tp}} w_{mi1},
\end{align}
from which non-Bayesian estimation of $\{ g_{mi1}, g_{mi2} \}$ is \emph{impossible}: the problem is unidentifiable (the Fisher information
is singular  and  the maximum-likelihood estimate is undefined). 
Other non-Bayesian estimates could be envisioned. 
For example, the following estimate (for the first UE) 
\begin{align}
\hat g_{mi1}   = \frac{1}{\sqrt{\Pp}}   y_{mi1}  = g_{mi1} + g_{mi2}  + \frac{1}{\sqrt{\Pp \tp}} w_{mi1},
\end{align}
represents the so-called ``least-squares'' estimate, but is substantially useless unless it is post-processed using additional a priori information about $g_{mi1}$~\cite{Bjornson2016c,Ozdogan2019}. 
In general, without the use of additional prior knowledge, estimates based on $\{y_{miq} \}$ are typically meaningless. The issue is of course that the AP sees the (reused) pilots superimposed, and without a priori information it has no way of telling which contribution to $y_{miq}$ originated from a specific UE.

\subsection{Proposed Method}
\label{subsec:proposed-method}
We propose a method to estimate the path losses $\{\beta_{mk} \}$ from pilot-contaminated observations that are useful irrespective of the channel fading distribution. We restrict our analysis herein to a system with at most two UEs that share the same pilot sequence. \textcolor{black}{(Such a system can multiplex simultaneously at most $2\tp$ UEs per coherence block.)}
The proposed approach works as follows:
\begin{enumerate}
\item A UE transmits phase-rotated versions of the assigned pilot sequence over different coherence blocks. The phase shifts needed to generate these phase-rotated pilots are not drawn uniformly at random as in~\cite{Kocharlakota2019}, but according to a pre-determined function known at the APs and UEs.
\item AP $m$ estimates the path losses towards the UEs through maximum-likelihood (ML) under a suitable assumption on $\{g_{mik} \}$. Let $L_{miq}$ be the contribution to the logarithm of the likelihood function (i.e., the logarithm of the joint probability distribution of the observed sample), with respect to the measurement $|y_{miq} |^2$ in coherence block $i$ for pilot $q$. These contributions are summed up over the coherence blocks to find the ML estimates of $\{\beta_{mk} \}$.
\item The APs estimate the channel responses by using MMSE, and decode the data using the so-obtained path loss estimates.
\item The APs de-rotate the channel estimates by using the pre-determined phase-rotation sequences.
\end{enumerate}
Without loss of generality, we assume that the first UE (UE 1) is assigned the pilot sequence with index $q$ and no pilot phase shift, for all the $I$ coherence blocks.   
Conversely, UE 2 applies $I$ phase rotations to its own assigned pilot $q$, according to a pre-determined function known to all APs, resulting in $I$ unique phase-shifted pilot sequences. These $I$ phase-rotated pilots are then transmitted in $I$ consecutive coherence blocks. The transmitted pilot from UE 2 in coherence block $i$ is given by $\sqrt{\Pp \tp}\bphi_q e^{j\varphi_i}$, where $\varphi_i \in [ -\pi, \pi ]$, and the phase shifting function is defined as 
\begin{equation} \label{eq:phase-rotation}
\varphi_i = \frac{(2i-1)\pi}{I} - \pi, \qquad i = 1,\ldots,I.
\end{equation}
This choice guarantees to draw $I$ phase shifts spread out over the unit circle.
The purpose of applying these structured phase rotations to the transmitted pilot sequence is to approximately de-correlate the channels of different UEs in different coherence blocks. This de-correlation facilitates the path loss estimation. In scenarios where the UEs' channels are uncorrelated (e.g., independent Rayleigh fading), phase-rotated pilots do not introduce significant benefits. Conversely, in line-of-sight (LoS), channels are highly correlated over multiple coherence blocks, and phase-rotated pilots are essential.

A similar idea was proposed in~\cite{Kocharlakota2019}, assuming only correlated Rayleigh fading channels, and consisting in UEs sending phase-rotated pilots, whose phase shifts are uniformly generated at random. 
Unlike in~\cite{Kocharlakota2019}, in our scheme the APs self-learn the path losses in the uplink training, thus requiring no additional pilot resources.

\subsection{Analysis for the LoS Channel}
\label{sec:LoS}
In LoS, we have $  |g_{m1k} |= \cdots = |g_{mIk}| = \sqrt{\beta_{mk}}$ and the phase of $g_{mik}$ is either constant or varies linearly with $i$. Assuming that two UEs are assigned the same pilot with index $q$ for $I$ coherence blocks, but transmit the pilots according to the proposed scheme, the de-spread pilot signal at AP $m$ in coherence interval $i$ is given by
\begin{equation}
y_{miq} = \sqrt{\Pp}( g_{mi1} + e^{j\varphi_{i}} g_{mi2}) + \frac{w_{miq}}{\sqrt{\tp}},
\end{equation} 
where $\varphi_i$ is given in~\eqref{eq:phase-rotation}.
To estimate the path losses, AP $m$ focuses on $|y_{miq}|^2$. An effective statistical distribution of the squared magnitude of the de-spread pilot signal, denoted by $Y$, can be obtained in closed form (parametrized by $\beta_{m1}$ and $\beta_{m2}$) as follows. For the sake of brevity, we let $\Pp=\tp=1$. Hence, $Y$ is given by  
\begin{align} \label{eq:y_collision}
Y = | g_{mi1} + e^{j\hat{\varphi}_i} g_{mi2} + w_{miq} |^2,
\end{align}
where $\!\hat{\varphi}_i\!\in\!\mathcal{U}[-\pi,\!\pi]$. Note that, $Y$ has the same distribution as 
\begin{equation}
Y' \triangleq \left| | \sqrt{  \beta_{m1}} + w_{miq}  | + e^{j\varphi'_i}  \sqrt{\beta_{m2}}  \right|^2,
\end{equation}
where $\varphi'_i \sim \mathcal{U}[0,\pi]$ by symmetry.
Let $A \triangleq \left| \sqrt{\beta_{m1}} + w_{miq}  \right|$, and $b \triangleq\sqrt{\beta_{m2}}$.
Conditioned on $w_{miq}$ (that enters only through $A$), the cumulative distribution function (cdf) is  
\begin{align}
& P(Y'\le t | A = a) = P( |a+be^{j\varphi'_i}|^2 \le t | A = a ) \nonumber  \\
 & \; = P( (a + b\cos(\varphi'_i))^2 + b^2 \sin^2(\varphi'_i) \le t | A = a) \nonumber \\
 & \; = P( a^2 + b^2 + 2a b \cos(\varphi'_i) \le t | A = a) \nonumber \\
 & \; = P \left( \cos(\varphi'_i) \le \frac{t-a^2-b^2}{2a b} \Bigg|~A = a \right) \nonumber \\ 
 & \; = \begin{cases}
1, & \sqrt{t} > a  + b, \\
0, & \sqrt{t} < |a-b|, \\
P\left( \varphi'_i \ge \arccos \left(\frac{t-a^2-b^2}{2a b} \right) \Big| A = a \right), & \mbox{otherwise},
\end{cases}
\end{align}
where in turn,
\begin{align}
& P\! \left(\! \varphi'_i \ge \arccos \left( \frac{t-a^2-b^2}{2a b} \right) \Bigg| A = a \! \right) \! \nonumber \\
& \quad = \! \frac{1}{\pi} \! \int_{\arccos \left( \frac{t-a^2-b^2}{2a b} \right)}^\pi d\varphi'_i \! =\! 1\!-\!\frac{1}{\pi} \arccos \left( \frac{t\!-\!a^2\!-\!b^2}{2a b} \right). 
\end{align}
Hence, the probability density function (pdf) of $Y'$ conditioned on $A$ is given by
\begin{align} \label{eq:pdf_collision}
&p_{Y'|A}(t~|~a) = \frac{\partial}{\partial t} P(Y'\le t|A = a) \nonumber \\
& \quad =
\begin{cases}
0, & t \in \mathcal{T}, \\
- \frac{1}{\pi} \frac{\partial}{\partial t}  \arccos \left( \frac{t-a^2-b^2}{2a b} \right) , & \mbox{otherwise,}
\end{cases}
\nonumber \\
& \quad = 
\begin{cases}
0, & t \in \mathcal{T}, \\
  \dfrac{1}{2 \pi a b \sqrt{ 1- \left(\frac{t-a ^2-b^2 }{2a  b }\right)^2 }}, & \mbox{otherwise,}
  \end{cases}
\end{align}
where $\mathcal{T}\!=\!(-\infty,|a\!-\!b|)\cup(a\!+\!b,+\infty)$.
But $A$ has a Ricean distribution with parameters $\sqrt{\beta_{m1}}$ and $1/\sqrt{2}$.\footnote{$R~\!\sim~\!\mathrm{Rice}(\nu,\sigma)$ if $R\!=\!\sqrt{X^2\!+\!Y^2}$, where $X~\!\sim~\!\mathcal{N}(\nu \cos{\theta},\sigma^2)$ and $Y~\!\sim~\!\mathcal{N}(\nu\sin{\theta},\sigma^2)$ are statistically independent normal random variables, and $\theta$ is any real number. Since $w_{miq}~\!\sim~\!\mathcal{CN}(0,1)$ and $\beta_{m1}$ is a constant, we have $\Re(\sqrt{\beta_{m1}} + w_{miq})~\!\sim~\!\mathcal{N}(\sqrt{\beta_{m1}},1/2)$, and $\Im(\sqrt{\beta_{m1}} + w_{miq})~\!\sim~\!\mathcal{N}(0,1/2)$. This gives $| \sqrt{\beta_{m1}} + w_{miq} |~\!\sim~\!\mathrm{Rice}\left(\sqrt{\beta_{m1}}, 1/\sqrt{2}\right)$.} So the unconditional pdf of $Y'$, parametrized by $\beta_{m1}$ and $\beta_{m2}$, is well approximated by
\begin{align} \label{eq:conditional-pdf-approx}
p_{Y'}(y'; \beta_{m1},\beta_{m2})  & = \int p_{Y'|A}(y'|a)~p_A(a)~d a \nonumber \\
& \approx \sum_j  p_{Y'|A}(y'|a_j)~ p_A(a_j)~\Delta_{a_j},
\end{align}
for a handful of points $\left\{ a_j \right\}$ appropriately selected from the support of the following pdf:
\begin{align}
p_A(a) = 2 a~e^{-(a^{2}+\beta_{m1})}I_{0}\left(2 a \sqrt{\beta_{m1}}\right),
\end{align}
where $\Delta_{a_j} = a_j - a_{j-1}$, and $I_0\left(2a\sqrt{\beta_{m1}}\right)$ is the $0$-th order \textit{modified Bessel} function of the first kind given by
\begin{equation*}
I_{0}\left(2a\sqrt{\beta_{m1}}\right)\!=\! \sum _{j=0}^{\infty }{\frac {(a^2\beta_{m1})^{j}}{j!~\Gamma (j+1)}}.
\end{equation*}

Assembling the components together, the contribution to the logarithm of the likelihood function, with respect to the measurement $\left|y_{miq}\right|^2$ is given by 
\begin{align}
&L_{miq}(\beta_{m1},\beta_{m2}) = \log p_{Y'}(y'; \beta_{m1},\beta_{m2}) \label{eq:likelihood} \\ 
&\quad \approx \log{\sum_j  p_{Y'|A}(y'|a_j)~ p_A(a_j)~\Delta_{a_j}} \nonumber \\ 
&\quad = \log\sum_j \begin{cases}
0, & y' \in \mathcal{Y}_j, \\
\dfrac{\Delta_{a_j}e^{-(a_j^{2}+\beta_{m1})}I_{0}\left(2 a_j \sqrt{\beta_{m1}}\right)}{\pi b \sqrt{ 1- \left(\frac{y'-a_j^2-b^2 }{2a_j  b }\right)^2}}, & \mbox{otherwise,}
  \end{cases}
\end{align}
where $\mathcal{Y}_j\!=\!(-\infty,|a_j-b|)\cup(a_j+b,+\infty)$.
These contributions are then summed up over the coherence blocks in order to find the maximum-likelihood estimates of $\{\beta_{mk}\}$, as follows
\begin{equation}\label{eq:likelihood-estimates}
\hat{\beta}_{m1}, \hat{\beta}_{m2} = \argmax_{\beta_{m1},\beta_{m2}} \sum^I_{i=1} L_{miq}(\beta_{m1},\beta_{m2}).
\end{equation}
\textcolor{black}{The complexity of the proposed scheme scales linearly with the number of co-pilot UEs and coherence blocks. Note that, the estimation decouples over the APs and the pilot sequences.}

\section{Simulation Results}
\begin{figure}[!t]
\centering
\includegraphics[width=\linewidth]{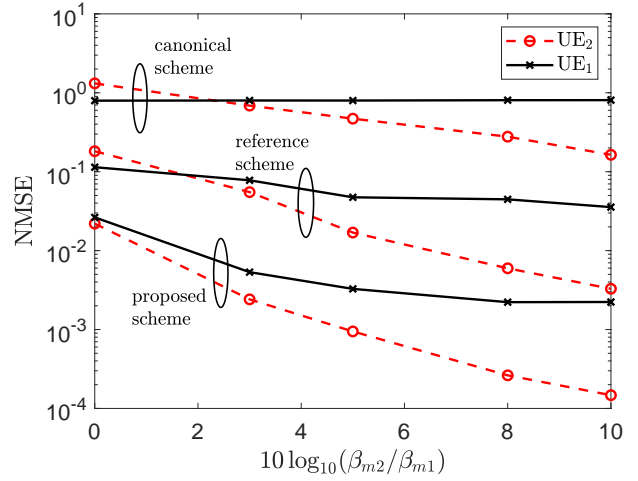}
\caption{NMSE of the path loss estimate in the case of structured phase-rotated pilot transmission (proposed scheme), psuedo-random phase-rotated pilots (reference scheme~\cite{Kocharlakota2019}), and non-phase-rotated pilots (canonical). $I~=~10$.}
\label{fig:fig1}
\end{figure}

We consider the system described in~\Secref{sec:two-user-system} wherein AP $m$ has to estimate, in $I$ coherence blocks, the path losses towards UE 1 and UE 2, which are assigned the same pilot sequence with index $q$. 
To evaluate the estimation performance, we measure the normalized mean square error (NMSE) of the path loss estimate, defined as
\begin{align}
\mathsf{NMSE}_k = \frac{1}{\beta_{mk}^2} {\EX{\left|\hat{\beta}_{mk}-\beta_{mk}\right|^2}}, \quad k = \{1, 2 \},
\end{align}
where $\{\hat{\beta}_{mk}\}$ are obtained by the ML detector in~\eqref{eq:likelihood-estimates} using a grid search. 
\Figref{fig:fig1} shows the NMSE of the path loss estimates with structured phase-rotated pilots (proposed), psuedo-random phase-rotated pilots~\cite{Kocharlakota2019} (reference), and canonical pilot transmission (i.e., no phase rotation is applied on the pilots).
We consider the LoS scenario described in~\Secref{sec:LoS}, and when generating the user channel, we set the path loss of UE 1 (referred to as \textit{reference} path loss) such that the corresponding large-scale uplink $\mathsf{SNR}_{m1}$, that is $\Pp \beta_{m1}$, is equal to 20 dB. Then, we let the path loss of UE 2 vary from 0 to 10 dB larger than the reference path loss. We also set $I=10$. The path loss estimates for all the schemes are obtained by using the detector in~\eqref{eq:likelihood-estimates}.
Firstly, we observe that phase-rotating the pilots in a deterministic fashion significantly improves the NMSE over the state-of-the-art. Note that, drawing phase shifts randomly, as in~\cite{Kocharlakota2019}, might result to poor estimates in the case of unfortunate shifts that de-correlate the channels only minimally, especially when few signal observations are available (i.e., $I$ is small). Secondly, the NMSE reduces as the path loss gap between UE 1 and UE 2 increases, as the detector is able to better differentiate the single contribution originated from each UE. Particularly, the NMSE of UE 2 substantially improves because of two factors: the excellent de-correlation among the observations from phase-rotating the pilots, and the higher $\mathsf{SNR}_{m2}$.

\begin{figure}[!t]
\centering
\includegraphics[width=\linewidth]{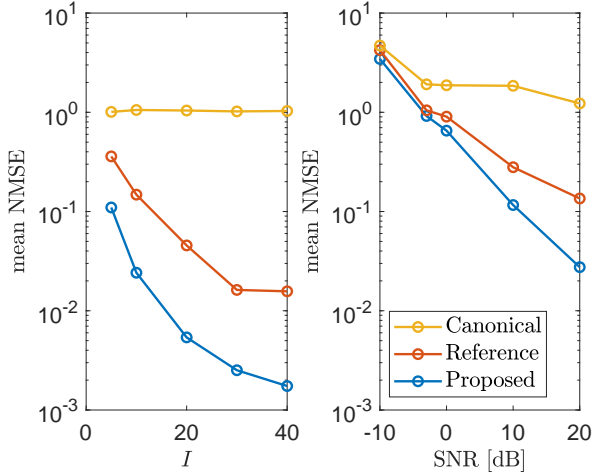}
\caption{Left: mean NMSE versus $I$, $\mathsf{SNR}_{m1} = \mathsf{SNR}_{m2} = 20$ dB. Right: mean NMSE versus large-scale SNR, $\beta_{m1}=\beta_{m2}$ and $I=10$. LoS scenario.}
\label{fig:fig2}
\vspace*{-2mm}
\end{figure}

In \Figref{fig:fig2}, we show how the mean NMSE varies with the length of the coherence block (left figure), and the large-scale SNR (right figure), in LoS scenario. In these simulations, we focus on the case when $\beta_{m1}=\beta_{m2}$, where the channel estimation is trickier, as observed in~\Figref{fig:fig1}.  
The left figure clearly shows that our proposed scheme needs very few coherence blocks to provide satisfactory mean NMSE. For example, our scheme employs just ten coherence blocks to obtain a mean NMSE slightly larger than $10^{-2}$, against about thirty coherence blocks used by the scheme in~\cite{Kocharlakota2019}. This result makes our scheme appealing in applications where channel conditions change very quickly, and/or with resource constraints.   
The performance of the canonical scheme does not vary with $I$ due to the low degree of uncorrelation of the channel observations. The right figure shows that all the schemes perform poorly when $\mathsf{SNR}_{m1}\!=\!\mathsf{SNR}_{m2}\!=\!-10$ dB, and the performance gain of our proposed scheme over the prior-art schemes increases as the large-scale SNR grows, resulting in less noisy channel observations. For instance, the NMSE provided by our proposed scheme at 3 dB SNR is achieved by the reference scheme at 10 dB SNR.


\vspace*{-1mm}
\section{Conclusion}
We proposed a path loss estimation method along with an accompanying phase-rotation pilot transmission scheme for the cell-free massive MIMO uplink, assuming line-of-sight channel model. The purpose of phase-rotating the pilot sequence in each coherence block, is to approximately de-correlate the channel observations at the APs.  
As a result, the proposed scheme can provide smaller NMSE at smaller SNR operation yet saving pilot resources compared to state-of-the-art schemes. These features are especially appealing in scenarios characterized by fast-changing blocking conditions, high carrier frequencies, low latency requirements, and bandwidth constraints.
We restricted our study to the case of systems with at most two co-pilot UEs. The general case with more co-pilot UEs may be included in a future work. \textcolor{black}{Another extension to this work may consist in devising a ML method that combines multiple statistics (e.g., $y_{miq}$ and $|y_{miq}|^2$) to directly estimate the channel responses.}

\bibliographystyle{IEEEtran}
\bibliography{IEEEabrv,refs}
\end{document}